\documentclass[10pt]{article}
\usepackage{nfam2025_workshop}
\usepackage{times}

\expandafter\let\csname equation*\endcsname\relax
\expandafter\let\csname endequation*\endcsname\relax
\usepackage{amsmath}
\usepackage{amssymb}
\usepackage{graphicx}
\usepackage{amsfonts}
\usepackage{xcolor}
\usepackage{comment}
\usepackage{natbib}

\usepackage{algorithm,algpseudocode}

\DeclareMathOperator{\bx}{\mathbf{x}}
\DeclareMathOperator{\bz}{\mathbf{z}}

\DeclareMathOperator{\bxi}{\boldsymbol{\xi}}
\DeclareMathOperator{\ve}{\varepsilon}

\title{The Capacity of Modern Hopfield Networks under the Data Manifold Hypothesis}

\author{%
Beatrice Achilli$^{1}$ \And
Luca Ambrogioni$^{2}$ \And
Carlo Lucibello$^{1}$ \And  
Marc Mézard$^{1}$ \And 
Enrico Ventura$^{1}$ \\ \\[1ex]
  $^{1}$Department of Computing Sciences, BIDSA, Bocconi University, Milan, MI 20100, Italy.\\
  $^{2}$Donders Institute for Brain, Cognition and Behaviour, Radboud University, \\6500 HD Nijmegen, The Netherlands.\\[1ex]
}

\iclrfinalcopy

\begin{document}

\maketitle

\begin{abstract}
We generalize the computation of the capacity of exponential Hopfield model from \cite{lucibello2023exponential} to more generic pattern ensembles, including binary patterns and patterns generated from a hidden manifold model. 
\end{abstract}

\section{Introduction}

Since its introduction in 1982, the Hopfield Model~\citep{hopfield1982neural} has been the standard model for associative memory. The retrieval of a full memory from partial information was shown to be possible for this model up to a critical capacity, which scales linearly with the system size ~\citep{Amit1985a}. 
Hopfield-like models with super-linear capacity have been investigated for a long time \cite{Gardner_1987}, and they achieved renewed popularity in recent years. In particular, \citet{krotov2016dense} discuss generalizations with polynomial capacity, while \citet{demircigil2017model} and \citet{ramsauer2020hopfield} introduce exponential capacity networks. In particular, these exponential models have attracted substantial interest in the past few years, partially due to their connections with transformer attention blocks~\citep{ramsauer2020hopfield, hoover2023energy} and generative diffusion models~\citep{ambrogioni_search_2023, hoover_memory_2023, ambrogioni2024thermo, biroli_dynamical_2024}. We will refer to the exponential capacity model of \cite{ramsauer2020hopfield} as to Modern Hopfield Network. In Ref. \citep{lucibello2023exponential}, the authors proposed the use of an auxiliary Random Energy Model (REM) to study the retrieval transition in MHNs with a signal-to-noise argument. In this paper, we further develop this approach and extend it to more general pattern ensembles, beyond the rotationally invariant ones they considered. 
In particular, we study structured patterns, generated according to the Hidden Manifold Model~\citep{goldt2020modeling} with non-linear activations. We use the replica formalism to study the auxiliary REM associated to the problem and characterize the critical retrieval threshold as a function of the latent dimensionality of the patterns.

\section{The Random Energy Model formalism}
\label{sec:REM}
MHNs are analogous to physical systems with randomly sampled energy levels, where each level corresponds to a pattern to be memorized \citep{lucibello2023exponential}. In statistical physics, the Random Energy Model is a well-known analytically tractable disordered system whose thermodynamic limit can be studied both either large deviation theory or the replica method \citep{Derrida1981,mezard2009information,fedrigo2007large}. In this section, we summarize the probabilistic tools needed to solve a generic REM. Let us consider $P=e^{\alpha N}$  i.i.d. energy levels $\ve^\mu \sim p(\ve \,|\, \omega)$, where $\alpha$ is a positive real number and we extend the typical REM setting allowing for a common source of quenched disorder $\omega \sim p_\omega$. In machine learning terms $\omega$ represents a common set of latent variables that influence the sampling of the observable patterns, or data points. 

We are interested in the thermodynamic properties of the system for a large number of energy levels, that is the large $N$ and $P$ limit. These properties can be obtained from the asymptotic free energy of the system at an inverse temperature $\lambda$, defined by
\begin{equation}
\phi_\alpha(\lambda) = \lim_{N\to\infty}  \frac{1}{\lambda N} \mathbb{E}\log \sum_\mu e^{\lambda N\ve^\mu}. 
\end{equation}
Notice that we adopt a convention where the energy has the opposite sign with respect to the usual physics convention.
In order to study the thermodynamic limit, it is convenient to introduce the cumulant generating function and its Legendre transform:
\begin{align}
\zeta(\lambda)&=\lim_{N\to\infty}\frac{1}{N}\mathbb{E}_\omega\log\mathbb{E}_{\ve|\omega}\, e^{\lambda N \ve},\label{eq:zeta}\\
s(\ve)&=\sup_{\lambda}\ \ve\lambda-\zeta(\lambda).\label{eq:s}
\end{align}
The generating function $\zeta(\lambda)$ is more tractable than $\phi_\alpha(\lambda)$ since the expectation with respect to $\varepsilon$ is moved inside the logarithm, where it can often be computed explicitly. 

The quantity $\Sigma_\alpha(\epsilon)=\alpha-s(\ve)$ corresponds to the annealed entropy of the system \citep{mezard2009information}. The annealed entropy corresponds to the correct (quenched) one, as long as it is positive. 
Let us define the condensation energy $\ve_{*}(\alpha)$ as the
largest root of $\Sigma_\alpha(\ve_{*})=0$. We have $\Sigma(\ve) <0$ for $\ve > \ve^*(\alpha)$. Let us also call $\tilde{\ve}(\lambda)=\zeta'(\lambda)$ the stationary energy.

For given $\lambda$ and $\alpha$, if $\Sigma(\tilde{\ve}(\lambda))=\alpha-s(\tilde{\ve}(\lambda))=\alpha+\zeta(\lambda)-\lambda\zeta'(\lambda)$ is larger than zero, the annealed description is correct, and an exponential number of energy levels with typical value $\tilde{\ve}(\lambda)$ dominate the partition function. This is called the \emph{uncondensed phase}.

Increasing $\lambda$, we select higher energy levels, until we reach the \emph{condensation transition} where the entropy becomes zero, and will stay zero also for larger values of $\lambda$. The partition function is dominated by the ground state $\ve^*(\alpha)$ in this region. Therefore we have what is called a \emph{condensed phase} for $\lambda > \lambda^*(\alpha)$, with $\lambda^*$ defined as the solution of:
\begin{equation}
  \alpha+\zeta(\lambda_{*})-\lambda_{*}\zeta'(\lambda_{*})=0.  
\end{equation}
Finally, the free energy is given by
\begin{equation}
\phi_{\alpha}(\lambda)=\begin{cases}
\frac{\alpha+\zeta(\lambda)}{\lambda} & \lambda<\lambda_{*}(\alpha),\\
\ve_{*}(\alpha) & \lambda\geq\lambda_{*}(\alpha).
\end{cases}\label{eq:phi-rem}
\end{equation}

\section{Modern Hopfield Networks}
\label{sec:hopfield}

Here we review and extend the formalism introduced in \cite{lucibello2023exponential} to compute the exact asymptotic thresholds for the capacity of the MHN introduced in \cite{Ramsauer2020}. 
While the study in \cite{lucibello2023exponential} was limited to rotational invariant pattern ensembles, we generalize it to a much larger class of ensembles for which the free energy of an auxiliary REM has the commonly found self-averaging property \citep{MPV}. 
We will apply the REM formalism to the computation of the capacity in presence of binary patterns (Section \ref{sec:binary}) and then to  patterns generated to live on a latent manifold (Section \ref{sec:manif}). Eventually, we consider the infinitive data-points scenario and prove that a linear manifold becomes the only attractor in this case.

We consider the generalization of the Hopfield model proposed in \cite{Ramsauer2020}. Given $P$ patterns 
$\bxi^\mu\in\mathbb{R}^N$ and continuous configurations $x\in\mathbb{R}^N$, the energy of the model is defined as
\begin{equation}
\label{eq:ene}
E(\mathbf{x})=-\frac{1}{\lambda}\log\sum_{\mu=1}^P e^{\lambda\bx\cdot\bxi^{\mu}}+\frac{1}{2}\lVert\bx\rVert^{2},
\end{equation}
where $\cdot$ denotes scalar product and $\lambda$ plays the role of a fictitious inverse temperature. We study the retrieval of a a generic pattern, say $\bxi^{1}$, by analyzing under which conditions the energy has a local minimum in correspondence of $\bxi^1$, in particular in the limit of large systems size $N$ and i.i.d. patterns. 
We shall assume that the components of $\bx$ and of $\bxi^\mu$ are of order 1, so that the energy scales linearly with $N$ at large N.

\subsection{The signal-to-noise argument}
\label{sec:stn}
An exact capacity threshold can be obtained using a simple signal-to-noise argument. The idea is that, if we are evaluating the energy close to a given pattern $\bxi^{1}$, all the other patterns can be interpreted as `unrelated noise' that interferes with the retrieval of the pattern. One can therefore split the energy into a signal and a noise contribution and approximate to exponential precision the energy density at large $N$ by
\begin{equation}
\frac{1}{N} E(\bx)\approx-\max\left(\frac{\bx\cdot\bxi^{1}}{N},\Phi(\bx)\right)+\frac{1}{2N}\lVert\bx\rVert^{2},
\end{equation}
where the noise function $\Phi$ is defined by
\begin{equation}
\Phi(\bx)=\frac{1}{\lambda N}\log\sum_{\mu\geq2}e^{\lambda\bx\cdot\bxi^{\mu}}.
\label{noisefunc}
\end{equation}
As we shall see, this function has a finite limit when $N\to+\infty$. The condition for retrieval is that the signal term in $\bx=\bxi^1$ dominates the noise term, that is $\lVert\bxi^{1}\rVert^2 > \Phi(\bxi^1)$. We consider the case in which we have an exponential number $P=e^{\alpha N}$, $\alpha >0$, of i.i.d. or conditionally i.i.d. patterns (as in the case of patterns produced by the hidden manifold model discussed later), normalized such that $\lim_{N\to\infty}\frac{1}{N}\mathbb{E}\lVert\bxi\rVert^2 \equiv r^2_\xi$ is finite. For reasonable pattern distributions, $\lVert\bxi^{1}\rVert^2/N$ and $\Phi(\bxi^1)$ will jointly concentrate on their expected value for large $N$. Therefore, depending on the values on the coupling strength $\lambda$ and the load $\alpha$, the retrieval condition $\lVert\bxi^{1}\rVert^2/N > \Phi(\bxi^1)$ is either almost always true or almost always false. The critical line $\alpha_1(\lambda)$ separating the retrieval phase  from the non-retrieval one when considering a uniformly chosen pattern, can be obtained solving the equation 
\begin{equation}
r^2_\xi = \lim_{N\to\infty} \frac{1}{N}\mathbb{E}\, \Phi(\bxi^1) = \phi_\alpha(\lambda). 
\label{eq:alpha1-cond}   
\end{equation}
Following \cite{lucibello2023exponential}, we now make the crucial observation that $\phi_\alpha(\lambda)$ corresponds to the asymptotic free energy of a REM~\citep{Derrida1981}, with distribution of energies induced by the distribution of patterns. This allows the computation of the free energy by using simple probabilistic techniques developed in this context.

The REM formalism introduced in Section~\ref{sec:REM} can be used to compute the single pattern retrieval threshold $\alpha_{1}(\lambda)$ obtained from Eq.~\eqref{eq:alpha1-cond}. Specifically, one can employ the expression of the free energy of a REM presented in Eq.~\eqref{eq:phi-rem}.

When the norm of the patterns is not bounded, we can have rare events in which the scalar product between $\bxi^1$ and another pattern with a very large norm can be dominating. In particular, when the typical norm is smaller than the maximum norm, there is a saturation effect in the capacity: even for very large $\lambda$, for $\alpha$ large enough you can always find a pattern that destabilizes the signal. This saturation of the single pattern retrieval threshold was already observed in \cite{lucibello2023exponential} in the case of Gaussian patterns.

\subsection{Modern Hopfield Networks with Binary patterns}
\label{sec:binary}

\begin{figure}[ht]
    \centering
    \includegraphics[width=0.5\linewidth]{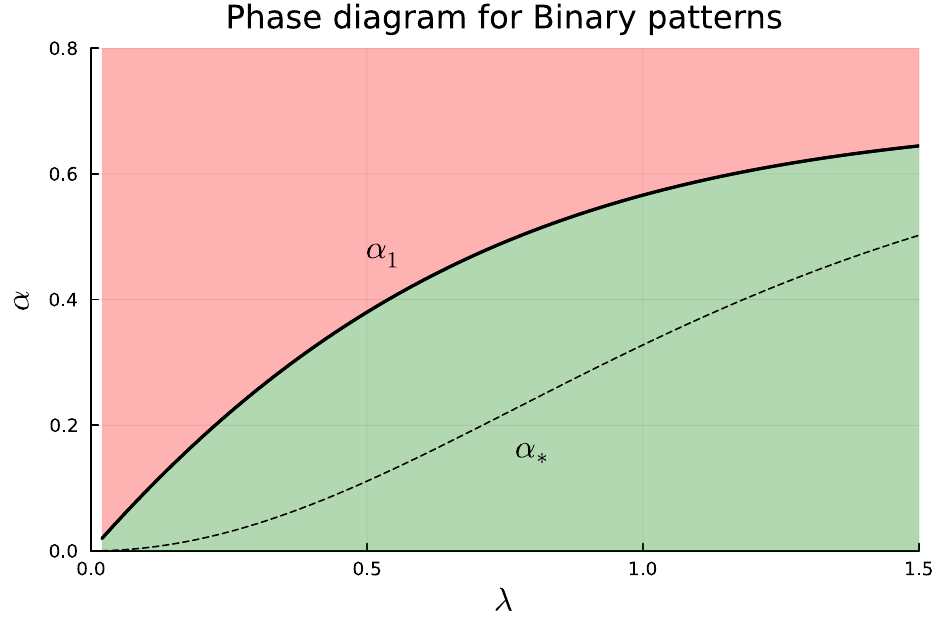}
     \caption{Phase diagram for binary patterns and variable. We also show the line $\alpha_{*}(\lambda)$ where the REM condensation occurs. }
     \label{fig:modern-hopfield2}
\end{figure}

As a first application, let us evaluate the retrieval capacity $\alpha_1(\lambda)$ from Eq.~\eqref{eq:alpha1-cond} in the case of binary pattern $\bxi^{\mu}\sim \mathrm{Unif}(\{-1,+1\}^{N})$. 
In order to compute $\mathbb{E}\Phi(\bxi^1) / N$, we first compute the generating and the rate function from Eqs.~\eqref{eq:zeta} and ~\eqref{eq:s}:  
\begin{align}
\zeta_{bin}(\lambda)&=\lim_{N\to\infty}\frac{1}{N}\mathbb{E}_{\bxi_1}\log\mathbb{E}_{\bxi_2}\, e^{\lambda \bxi^1\cdot \bxi^2} = \log\cosh(\lambda),\\
s_{bin}(\epsilon)  &= \frac{1+\epsilon}{2} \log(1+\epsilon)+ \frac{1-\epsilon}{2}\log(1-\epsilon).
\end{align}
Next, we compute numerically $\epsilon_{*}(\alpha)$ solving $s_{bin}(\epsilon)=\alpha$ for $\ve$,
and set $\lambda_{*}(\alpha)=s_{bin}'(\epsilon_{*}(\alpha))$. At this point, we can compute the REM free energy $\phi_{\alpha}(\lambda)$ from Eq.~\eqref{eq:phi-rem} and we can find the single pattern retrieval threshold $\alpha_{1}(\lambda)$
as the $\alpha$ for which $\phi_{\alpha}(\lambda)=1$. The corresponding phase diagram is shown in Fig.~\ref{fig:modern-hopfield2}.

\subsection{Patterns from the Hidden Manifold Model}
\label{sec:manif}
\begin{figure}[ht]
    \centering
    \includegraphics[width=0.32\linewidth]{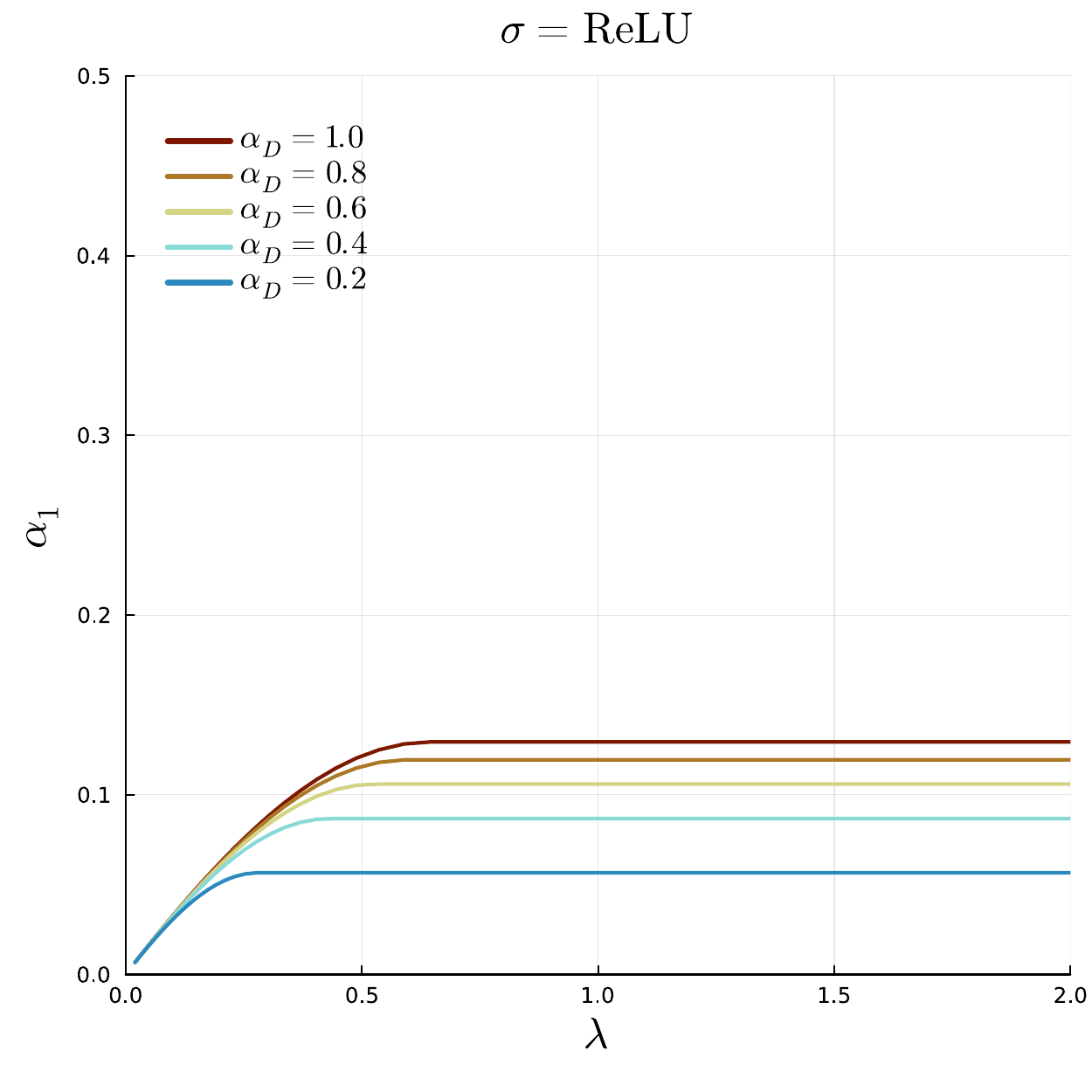}
    \includegraphics[width=0.32\linewidth]{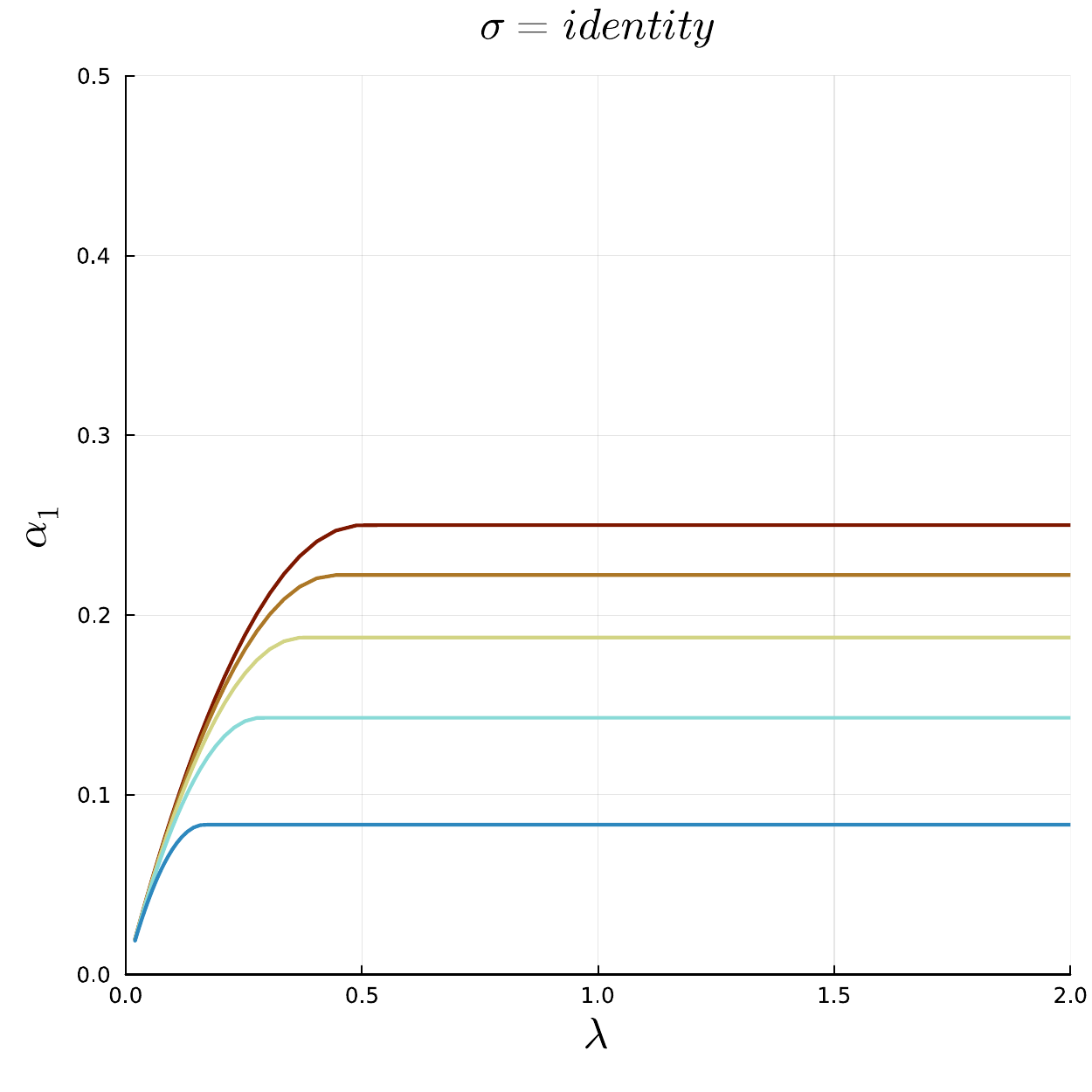}
    \includegraphics[width=0.32\linewidth]{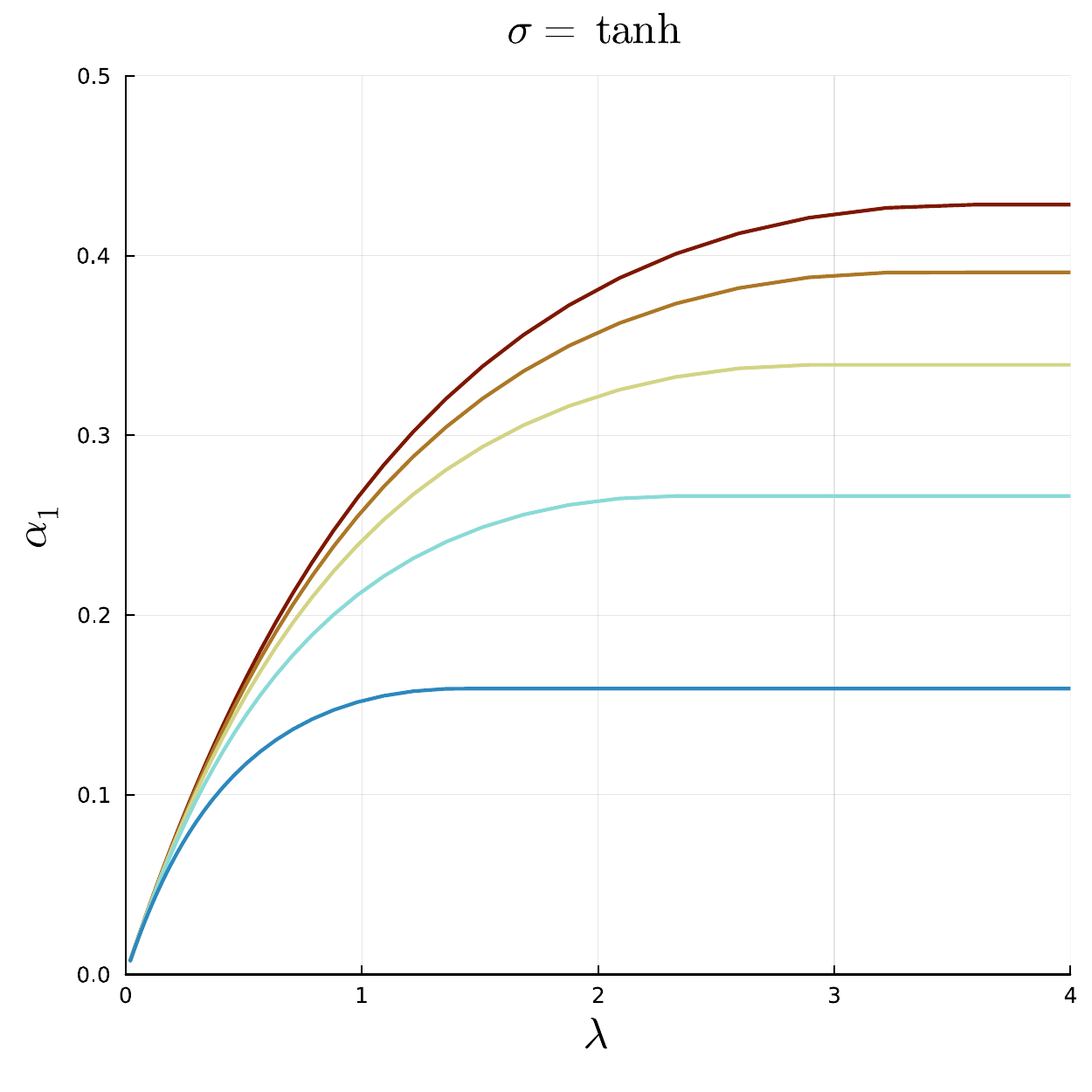}
    
    \caption{The retrieval lines $\alpha_{1}(\lambda)$ for patterns from an Hidden Manifold Model with $\mathrm{ReLU}$ (\textbf{left}), linear (\textbf{center}) and $\tanh$ (\textbf{right}) activations and different aspect ratios $\alpha_{D}=D/N$.}
    \label{fig:modern-hopfield}
\end{figure}

Let us now consider patterns generated by a Hidden Manifold Model  (HMM). The HMM is a simple synthetic generative process displaying the idea of the data manifold hypothesis, where data lies on $D$-dimensional sub-manifold of the ambient $N$-dimensional space. This generative process has been introduced and investigated in \cite{goldt2020modeling,goldt2022gaussian,gerace2020generalisation} in the supervised learning context and studied in \cite{negri_storage_2023} in the standard Hopfield case. Patterns are generated as $\bxi^{\mu}=\sigma\left(\frac{1}{\sqrt{D}}F \bz^{\mu}\right)$, where the latent variables $\bz^\mu$ are Gaussian, $\bz^{\mu}\sim\mathcal{N}(0,I_{D})$, $\sigma$ is an element-wise
non-linearity, and $F\in\mathbb{R}^{N\times D}$ is a random matrix with i.i.d. standard Gaussian entries. We define
$\alpha_D = D/N$, and assume $D,N\to+\infty$ with $\alpha_D$ having a finite limit. 
The REM generating function \eqref{eq:zeta} can be written as 
\begin{equation}
\zeta_{HMM}(\lambda)=\lim_{N\to\infty}\frac{1}{N}\mathbb{E}_{F,
\bz^{1}}\log\mathbb{E}_{
\bz^{2}}\,e^{\lambda\,\sigma\left(\frac{1}{\sqrt{D}}F\bz^1\right)\,\cdot\,\sigma\left(\frac{1}{\sqrt{D}}F\bz^2\right)}
\end{equation}
We leverage the replica method to perform the computation, that is for a given integer $n>0$ we write the replicated partition function 
\begin{equation}
\mathbb{E}Z_{HMM}^n=\mathbb{E}_{F,
\bz,\bz^{1:n}}\,e^{\lambda\sum_{a=1}^n\sigma\left(\frac{1}{\sqrt{D}}F\bz\right)\,\cdot\,\sigma\left(\frac{1}{\sqrt{D}}F\bz^a\right)},
\end{equation}
we perform saddle point evaluation within a Replica Symmetric (RS) ansatz, and we obtain by analytical continuation the result for $\zeta_{HMM}(\lambda)=\lim_{n\to0}\lim_{N\to\infty}\frac{1}{n N}\log\mathbb{E}Z_{HMM}^n$. The details of the computation are provided in \ref{appendix:replicaHMM}. The result is
\begin{equation}
\zeta_{HMM}(\lambda)=-\alpha_{D}m\hat{m}-\frac{\alpha_{D}}{2}(q_{d}\hat{q}_{d}-q_{0}\hat{q}_{0})+\alpha_D G_{S}+G_{E},
\label{eq:zetaHMM}
\end{equation}
with
\begin{align}
G_{S} & =-\frac{1}{2}\log\left(1-\hat{q}_{d}+\hat{q}_{0}\right)+\frac{1}{2}\frac{\hat{m}^{2}+\hat{q}_{0}}{1-\hat{q}_{d}+\hat{q}_{0}},\\
G_{E} & =\int Dt\;\int Du^{0}\,\log\int Du \  e^{\lambda\sigma\left(u^{0}\right)\,\sigma\left(\sqrt{q_d-q_0}u+mu^{0}-\sqrt{q_{0}-m^{2}}t\right)}.
\end{align}
Here $\int Dx$ denotes standard Gaussian integration. Eq.~\eqref{eq:zetaHMM} has to be evaluated on its stationary point with respect to the auxiliary parameters $q_d,\hat{q}_{d},q_0,\hat{q}_{0}$.
Solving numerically the saddle point equations and then using the usual REM formalism we can obtain the single
pattern retrieval threshold $\alpha_{1}(\lambda)$. Results are shown for patterns from a linear manilfold, ReLU and tanh activations in Fig.~\ref{fig:modern-hopfield}. 
The capacity decreases as the latent dimension $D$ shrinks. Notice that this is not straightforward, since the expected distance between the patterns does not depend on $D$. 
Another phenomena is the saturation effect that we can observe in all the three studied cases, though for $\sigma= \tanh$ it takes place at larger $\lambda$ values. This was anticipated in Section~\ref{sec:stn}: if the typical norm of the patterns is smaller than the maximum norm, even at large $\lambda$, for $\alpha$ big enough we will find patterns with large scalar product in the noise term. 

\section{Generalization in Modern Hopfield Networks}
\label{sec:gene}

We hereby consider data generated on a linear manifold defined by a feature matrix $F$. In the previous sections we have proved the existence of a maximum amount of patterns that can be retrieved as local minima of the energy function defined in Eq.~\eqref{eq:ene}. We will now show that, by increasing the number of patterns, we can bring the model closer to the condition of having the entire latent manifold as the only attractor in the landscape. We call this specific scenario \textit{generalization} limit of the MHN.    
Consider data generated on a linear hidden manifold as $\sigma(F\bz) = F\bz$, where we have re-absorbed the factor $1/\sqrt{D}$ into $F$ for simplicity of the notation.  Then we can compute the energy gradient-descent dynamics driving the memory retrieval as
\begin{equation}
    \mathbf{x}^{(t+1)} = \mathbf{x}^{(t)} - \nabla_{\bx} E(\mathbf{x}^{(t)}),
\end{equation}
Let us rewrite the energy function of the MHN as
\begin{align}
    E(\bx) &= -\frac{1}{\lambda}\log\int D\bz e^{\lambda\bx F\bz} + \frac{1}{2}\lVert \bx\rVert^2\\ 
&= -\frac{1}{\lambda}\log\left(e^{\frac{\lambda^2}{2}\bx^{\top}F F^{\top}\bx}\right)+ \frac{1}{2}\lVert \bx\rVert^2\\ 
&=\frac{1}{2}\bx^{\top}\left(\mathbf{1}_N - \lambda F F^{\top} \right)\bx,
\end{align}
By choosing $\lambda = 1$
we obtain the following updating rule for the dynamics
\begin{equation}
    \mathbf{x}^{(t+1)} =F F^{\top}\mathbf{x}^{(t)},
\end{equation}
which projects any configuration $\mathbf{x}^{(t)}$ on the data-manifold in one single iteration. The entire manifold is now  attractive for the gradient descent dynamics.  

\section{Conclusions}
In this paper, we extended the formalism of the Random Energy Model applied to study the retrieval transition in dense associative memory to more general pattern ensembles. In particular, we analyzed a simple case of structured patterns, the Hidden Manifold Model, considering the effect that the latent dimension and the non-linearity have on the retrieval of a pattern. Even if the typical distance between patterns does not depend on the hidden dimension, we found that it affects the retrieval threshold, which decreases with a smaller latent dimension. It would be interesting to extend this study to what happens beyond the retrieval threshold. For instance, one possible research direction is to analyze the stability of mixtures of patterns as in \cite{kalaj2024randomfeatureshopfield}, to see how generalization extends to the exponential regime.
We finally stress the importance of studying MHNs because of their formal equivalence with empirical-score-driven Diffusion Models (DMs), state-of-the-art generative models \citep{ventura2024spectral, achilli2024losing, biroli_dynamical_2024,ambrogioni2024thermo}. Specifically, the energy function of a MHN with spherical patterns (as treated in \cite{lucibello2023exponential}) coincides with the time-dependent potential of a DM by substituting $1/\lambda$ with the diffusion time $t$.  

\section*{Acknowledgements}
This publication benefited from European Union - Next Generation EU funds, component M4.C2, investment 1.1. - CUP J53D23001330001.

\bibliographystyle{plainnat}
\bibliography{main_workshop.bib}

\newpage
\appendix

\section{Replica calculation for the MHN capacity with HMM patterns}

\label{appendix:replicaHMM}

Setting $n=0$ fixes the reference-only overlaps, i.e. $q_{00}=1$
and $\hat{q}_{00}=0$.

The RS ansatz is
\begin{equation}
q_{ab}=\left(\begin{array}{cccc}
1 & m & \ldots & m\\
m & q_{d} &  & q_{0}\\
\vdots &  & \ddots\\
m & q_{0} &  & q_{d}
\end{array}\right) ;\ \hat{q}_{ab}=\left(\begin{array}{cccc}
0 & \hat{m} & \ldots & \hat{m}\\
\hat{m} & \hat{q}_{d} &  & \hat{q}_{0}\\
\vdots &  & \ddots\\
\hat{m} & \hat{q}_{0} &  & \hat{q}_{d}
\end{array}\right)\label{eq:rs-ansatz}
\end{equation}
The entropic term is
\begin{align}
G_{S} & =\frac{1}{n}\log\int\prod_{a=0}^{n}\frac{dz^{a}}{\sqrt{2\pi}}\ e^{-\frac{1}{2}\sum_{a}\left(z^{a}\right)^{2}}e^{\frac{1}{2}\sum_{ab}\hat{q}_{ab}z^{a}z^{b}}\\
 &= \frac{1}{n}\log\int\frac{dz^{0}}{\sqrt{2\pi}}\prod_{a'=1}^{n}\frac{dz^{a'}}{\sqrt{2\pi}}\ e^{-\frac{1}{2}\left(z^{0}\right)^{2}+\hat{m}z^{0}\sum_{a'}z^{a'}}e^{-\frac{1}{2}\left(1-\left(\hat{q}_{d}-\hat{q}_{0}\right)\right)\sum_{a'}\left(z^{a'}\right)^{2}+\frac{1}{2}\hat{q}_{0}\left(\sum_{a'}z^{a'}\right)^{2}}\\
& = \frac{1}{n}\log\int\prod_{a'=1}^{n}\frac{dz^{a'}}{\sqrt{2\pi}}\ e^{\frac{1}{2}\hat{m}^{2}\left(\sum_{a'}z^{a'}\right)^{2}}e^{-\frac{1}{2}\left(1-\hat{q}_{d}+\hat{q}_{0}\right)\sum_{a'}\left(z^{a'}\right)^{2}+\frac{1}{2}\hat{q}_{0}\left(\sum_{a'}z^{a'}\right)^{2}}\\
 & =\frac{1}{n}\log\int Dt\;\int\prod_{a'=1}^{n}\frac{dz^{a'}}{\sqrt{2\pi}}\ e^{-\frac{1}{2}\left(1-\hat{q}_{d}+\hat{q}_{0}\right)\sum_{a'}\left(z^{a'}\right)^{2}}e^{\sqrt{\hat{m}^{2}+\hat{q}_{0}}t\sum_{a'}z^{a'}}\\
 & =\frac{1}{n}\log\int Dt\;\left(\frac{1}{\sqrt{1-\left(\hat{q}_{d}-\hat{q}_{0}\right)}}e^{\frac{1}{2}\frac{\left(\hat{m}^{2}+\hat{q}_{0}\right)}{\left(1-\hat{q}_{d}+\hat{q}_{0}\right)}t^{2}}\right)^{n}\end{align}
 so that
 \begin{align}
\lim_{n\to0} G_S= -\frac{1}{2}\log\left(1-\hat{q}_{d}+\hat{q}_{0}\right)+\frac{1}{2}\frac{\hat{m}^{2}+\hat{q}_{0}}{1-\hat{q}_{d}+\hat{q}_{0}}
\end{align}
while the energetic part is
\begin{align}
G_{E}= & \frac{1}{n}\log\int\prod_{a=0}^{n}\frac{du^{a}d\hat{u}^{a}}{2\pi}\ e^{\lambda\,\sum_{a'=1}^{n}\sigma\left(u^{0}\right)\,\sigma\left(u^{a'}\right)-\sum_{a}i\hat{u}^{a}u^{a}-\frac{1}{2}\sum_{ab}\hat{u}^{a}\hat{u}^{b}q_{ab}}\\
= & \frac{1}{n}\log\int\frac{du^{0}d\hat{u}^{0}}{2\pi}\prod_{a'=1}^{n}\frac{du^{a'}d\hat{u}^{a'}}{2\pi}\ e^{\lambda\,\sum_{a'=1}^{n}\sigma\left(u^{0}\right)\,\sigma\left(u^{a'}\right)-\sum_{a'}i\hat{u}^{a'}u^{a'}-i\hat{u}^{0}u^{0}}\nonumber \\
 & \times e^{-\frac{1}{2}\left(\hat{u}^{0}\right)^{2}-m\hat{u}^{0}\sum_{a'}\hat{u}^{a'}-\frac{1}{2}(q_{d}-q_{0})\sum_{a'}\left(\hat{u}^{a'}\right)^{2}-\frac{1}{2}q_{0}\left(\sum_{a'}\hat{u}^{a'}\right)^{2}}\\
= & \frac{1}{n}\log\int\frac{du^{0}}{\sqrt{2\pi}}\prod_{a'=1}^{n}\frac{du^{a'}d\hat{u}^{a'}}{2\pi}\ e^{\frac{1}{2}\left(m\sum_{a'}\hat{u}^{a'}+iu^{0}\right)^{2}}e^{\lambda\,\sum_{a'=1}^{n}\sigma\left(u^{0}\right)\,\sigma\left(u^{a'}\right)-\sum_{a'}i\hat{u}^{a'}u^{a'}}\nonumber \\
 & \times e^{-\frac{1}{2}(q_{d}-q_{0})\sum_{a'}\left(\hat{u}^{a'}\right)^{2}-\frac{1}{2}q_{0}\left(\sum_{a'}\hat{u}^{a'}\right)^{2}}\\
= & \frac{1}{n}\log\int Dt\;\int\frac{du^{0}}{\sqrt{2\pi}}\prod_{a'=1}^{n}\frac{du^{a'}d\hat{u}^{a'}}{2\pi}\ e^{-\frac{\left(u^{0}\right)^{2}}{2}}e^{\lambda\,\sum_{a'=1}^{n}\sigma\left(u^{0}\right)\,\sigma\left(u^{a'}\right)}\nonumber \\
 & \times e^{-i\sum_{a'}\hat{u}^{a'}\left(u^{a'}-mu_{0}+\sqrt{q_{0}-m^{2}}t\right)-\frac{1}{2}(q_{d}-q_{0})\sum_{a'}\left(\hat{u}^{a'}\right)^{2}}\\
= & \frac{1}{n}\log\int Dt\;\int Du^{0}\\
 & \times\left(\int\frac{du}{\sqrt{2\pi}}\ e^{\lambda\sigma\left(u^{0}\right)\,\sigma\left(u\right)}\frac{1}{\sqrt{q_{d}-q_{0}}}e^{-\frac{1}{2(q_{d}-q_{0})}\left(u-mu^{0}+\sqrt{q_{0}-m^{2}}t\right)^{2}}\right)^{n}\end{align}
 So that
 \begin{align}
\lim_{n\to0} G_E
 & =\int Dt\;\int Du^{0}\log\left(\int\frac{du}{\sqrt{2\pi(q_{d}-q_{0})}}\ e^{\lambda\sigma\left(u^{0}\right)\,\sigma\left(u\right)-\frac{1}{2(q_{d}-q_{0})}\left(u-mu^{0}+\sqrt{q_{0}-m^{2}}t\right)^{2}}\right).
\end{align}
We can then write
\begin{align}
\zeta_{\lambda}(q_{d},q_{0},m,\hat{q}_{d},\hat{q}_{0},\hat{m}) & =-\alpha_{D}m\hat{m}-\frac{\alpha_{D}}{2}(q_{d}\hat{q}_{d}-q_{0}\hat{q}_{0})-\frac{\alpha_{D}}{2}\log\left(1-\hat{q}_{d}+\hat{q}_{0}\right)+\frac{\alpha_{D}}{2}\frac{\hat{m}^{2}+\hat{q}_{0}}{1-\hat{q}_{d}+\hat{q}_{0}}\nonumber \\
 & +\int Dt\;\int Du^{0}\log\left(\int Du \  e^{\lambda\sigma\left(u^{0}\right)\,\sigma\left(\sqrt{q_d-q_0}u+mu^{0}-\sqrt{q_{0}-m^{2}}t\right)}\right)
\end{align}
and take its derivatives to obtain the saddle point equations
\begin{align}
\hat{q}_{d} & =\frac{2}{\alpha_{D}}\int Dt\;\int Du^{0}\int Du \  e^{\lambda\sigma\left(u^{0}\right)\,\sigma\left(\sqrt{q_d-q_0}u+mu^{0}-\sqrt{q_{0}-m^{2}}t\right)}\nonumber  \\
&\times\Bigg[\frac{\frac{\left(u-mu^{0}+\sqrt{q_{0}-m^{2}}t\right)^{2}-\left(q_{d}-q_{0}\right)}{2\left(q_{d}-q_{0}\right)^{2}}}{{\int Du \  e^{\lambda\sigma\left(u^{0}\right)\,\sigma\left(\sqrt{q_d-q_0}u+mu^{0}-\sqrt{q_{0}-m^{2}}t\right)}}}\Bigg]\\
\hat{q}_{0} & =-\frac{2}{\alpha_{D}}\int Dt\;\int Du^{0}\int Du \  e^{\lambda\sigma\left(u^{0}\right)\,\sigma\left(\sqrt{q_d-q_0}u+mu^{0}-\sqrt{q_{0}-m^{2}}t\right)}\nonumber \\
&\times \Bigg[\frac{\Bigg(\frac{\left(u-mu^{0}+\sqrt{q_{0}-m^{2}}t\right)^{2}}{2\left(q_{d}-q_{0}\right)^{2}}+\frac{\left(u-mu^{0}+\sqrt{q_{0}-m^{2}}t\right)\sqrt{q_{0}-m^{2}}t+1}{2\left(q_{d}-q_{0}\right)}\Bigg)}{{\int Du \  e^{\lambda\sigma\left(u^{0}\right)\,\sigma\left(\sqrt{q_d-q_0}u+mu^{0}-\sqrt{q_{0}-m^{2}}t\right)}}}\Bigg]
\\
\hat{m} & =\frac{1}{\alpha_{D}}\int Dt\;\int Du^{0}\int Du \  e^{\lambda\sigma\left(u^{0}\right)\,\sigma\left(\sqrt{q_d-q_0}u+mu^{0}-\sqrt{q_{0}-m^{2}}t\right)}\nonumber \\
&\times\Bigg[\frac{\frac{\left(u-mu^{0}+\sqrt{q_{0}-m^{2}}t\right)\left(u^{0}+mt(q_{0}-m^{2})^{-1/2}\right)}{\left(q_{d}-q_{0}\right)}}{\int Du \  e^{\lambda\sigma\left(u^{0}\right)\,\sigma\left(\sqrt{q_d-q_0}u+mu^{0}-\sqrt{q_{0}-m^{2}}t\right)}}\Bigg]\\
q_{d} & =\frac{1}{1-\hat{q}_{d}+\hat{q}_{0}}+\frac{\hat{m}^{2}+\hat{q}_{0}}{\left(1-\hat{q}_{d}+\hat{q}_{0}\right)^{2}}\\
q_{0} & =\frac{1}{1-\hat{q}_{d}+\hat{q}_{0}}+\frac{\hat{m}^{2}+\hat{q}_{d}-1}{\left(1-\hat{q}_{d}+\hat{q}_{0}\right)^{2}}\\
m & =\frac{\hat{m}}{1-\hat{q}_{d}+\hat{q}_{0}}
\end{align}
\end{document}